\documentclass[prl, showpacs, twocolumn]{revtex4}
\usepackage{graphicx}
\usepackage{amsmath}
\usepackage{amssymb}

\begin{document}

\title{Temporal decorrelation of collective oscillations in neural networks\\ with local inhibition and long-range excitation}

\author{Demian Battaglia}
\author{Nicolas Brunel}
\author{David Hansel}
\affiliation{Laboratoire de Neurophysique et Physiologie, Universit\'e Paris
Descartes, CNRS UMR 8119;\\45, rue des Saints-P\`eres, 75270 Paris Cedex 06,
France}

\begin{abstract}

We consider two neuronal networks coupled by long-range excitatory interactions. Oscillations in the gamma frequency band are generated within each network by local inhibition. 
When long-range excitation is weak, these oscillations phase-lock with a phase-shift dependent on the strength of local inhibition. Increasing the strength of long-range excitation induces a transition to chaos via period-doubling or quasi-periodic scenarios. In the chaotic regime oscillatory activity undergoes fast temporal decorrelation. The generality of these dynamical properties is assessed in firing-rate models as well as in large networks of conductance-based neurons.

\end{abstract}


\maketitle

Fast synchronous gamma rhythms 
\mbox{(30--100 Hz)} are observed in the neuronal activity of cortical areas \cite{gammabookBuzsaki, Gray89, Bragin95, Chrobak98}. Modeling  \cite{WangBuzsaki96, Traub97, Brunel99, WhittingtonKopell00} and experimental studies \cite{WhittingtonKopell00, Whittington95, BartosVida07} suggest that these oscillations can be generated within local networks of GABAergic inhibitory interneurons. 

Gamma oscillatory episodes lose temporal 
coherence in several tens of milliseconds  \cite{Kreiter96, Csicsvari03}. 
This decorrelation could be due to noisy extrinsic feed-forward inputs, 
but this would require substantial spatial 
correlations in their fluctuations on the scale of the circuits 
generating the rhythm. 
Another possibility is that damped crosscorrelograms 
arise because the network is in fact close to the onset 
of synchrony  \cite{Brunel99}. In this case
temporal decorrelation is a finite size effect.  
Here we explore an alternative mechanism in which the decoherence of 
gamma oscillations emerges as a collective phenomenon.

We study a model of a small piece of cortex (2--6 mm$^2$) which consists of two local 
networks representing two cortical columns \cite{Mountcastle97}. Each network 
comprises one inhibitory and one excitatory population of neurons. 
Oscillations are induced independently within each network by 
inhibitory to inhibitory interactions \cite{Brunel99, Roxin05}.
The two networks interact via their excitatory populations, as consistent
with anatomical evidences that excitation has a larger spread than inhibition \cite{Albus94, Bosking97}.  
We show below, in both firing-rate and conductance-based models, 
that when this long-range excitation 
is sufficiently strong, the activity of the whole system 
displays synchronous but highly irregular oscillations with fast 
temporal decorrelation. This decorrelation of the rhythmic activity 
is associated with the emergence of stable chaotic attractors.

We start by a simplified model in which the activity of each excitatory and inhibitory population is described by a firing-rate variable:
\begin{equation}\label{RateFull}
\begin{split}
\tau_{{}_E}\dot{m}^E_{1,2}(t) & = -m^E_{1,2}(t) + \Phi\left[h^E_{1,2} + S^{EE}\,m^E_{1,2}(t-D)\right.\\
&\left.\quad + S^{EI}\,m^I_{1,2}(t-D) + L^{EE}\,m^E_{2,1}(t-\bar{D}) \right]\\
\tau_{{}_I}\dot{m}^I_{1,2}(t) & = -m^I_{1,2}(t) + \Phi\left[ h^I_{1,2} + S^{II}\,m^I_{1,2}(t-D)\right.\\
&\left.\quad + S^{IE}\,m^E_{1,2}(t-D) + L^{IE}\,m^E_{2,1}(t-\bar{D}) \right]
\end{split}
\end{equation}  
where $m_i^\alpha$, $\alpha=E,I$ and $i=1,2$, are the activities of 
population $\alpha$ in the local network $i$ and $h_i^\alpha$ are 
external driving currents, constant in time. The interaction between populations 
$\alpha$ and $\beta$ within a local network is denoted by $S^{\alpha\beta}$ and 
the interaction between the two networks by $L^{\alpha E} > 0$ ($\alpha,\beta= E,I$).
The delays $D$ and $\bar D$ (in the intra- and inter-network interactions, respectively) represent synaptic and conduction delays. We choose a threshold-linear
transfer function $\Phi[x] = [x]_+ = x$ if $x>0$, 0 otherwise. For simplicity 
we take $\tau_{{}_E}=\tau_{{}_I}=1$, $h^E_{1,2} = h^I_{1,2} \equiv h_{ext}$, $S^{EI}=S^{II}$, 
$S^{EE}=S^{IE}$ and $L^{EE}=L^{IE}$. Hence, in any 
attractor of the dynamics $m_i^E=m^I_i\equiv m_i$. The equations~\eqref{RateFull} 
then reduce to:
\begin{equation}\label{RateEquation}
\begin{split}
\dot{m}_{1,2}(t) & = -m_{1,2}(t) + [ h_{ext} \\ &\quad + K_0\,m_{1,2}(t-D) + K_1\,m_{2,1}(t-\bar{D})]_+
\end{split}
\end{equation}
where $K_0 = S^{II} + S^{IE}$ and $K_1 = L^{IE} > 0$. Now the model describes two effective populations with a local (intra-population) interaction $K_0$ and a long-range (inter-population) interaction $K_1$. 

The analysis of  Eq.~\eqref{RateEquation} simplifies if we assume $D=\bar{D}$
as we do in most of the paper.  However, the results described below remain 
qualitatively valid for a broad range of values of $D$ and $\bar{D}$ even if 
$D \neq \bar{D}$ \cite{SuppMat}.

\begin{figure}
\begin{center}
\includegraphics[scale=0.377]{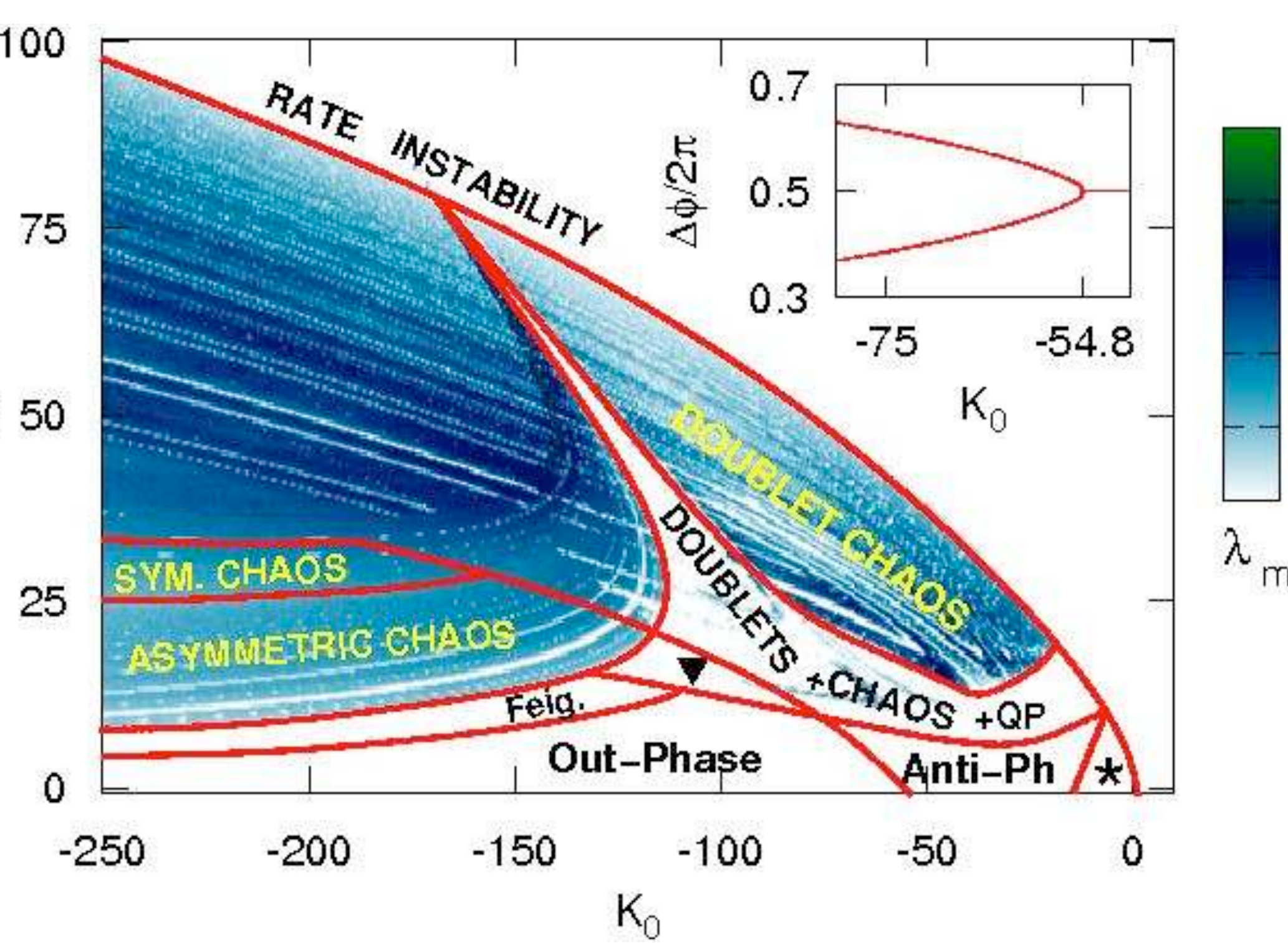}
\end{center}
\caption{Phase diagram of the two-population rate model ($D=\bar D = 0.1$). Background color indicates positive $\lambda_{max}$. Only the discussed bifurcation lines are plotted. (*) designs the stable homogeneous fixed-point region, ($\blacktriangledown$) a region of multistability not analyzed in the present paper. Inset: phase-shift in the weak coupling limit ($K_0^c\simeq -54.8$).}\label{fig:phasediagram}
\end{figure}
Let us first consider the dynamics of one isolated population ($K_1=0$).
For sufficiently strong local inhibition, at $K_0= K_0^{osc}(D)$, 
the fixed point, $m_{i}(t)=h_{ext}/(1-K_0)$, loses stability 
via a Hopf bifurcation.  For  $D \ll 1$ , 
$K_0^{osc}(D)\sim - \pi/(2D)$. The activity, $m_{osc}(t)$, of 
the population in the resulting oscillatory regime can be derived under certain 
conditions \cite{Roxin05, SuppMat}. It is also possible, under these same conditions, 
to compute the phase-response curve $Z(\phi_p)=\partial\phi/\partial h$, 
which quantifies the shift in the phase, $\phi$, of the population 
oscillation, following a small current perturbation 
$\delta h$ 
applied at phase $\phi_p$  \cite{Kuramoto, SuppMat}. 

When the populations are weakly coupled \mbox{($K_1\rightarrow 0^+$)}, 
the oscillations in their activity become phase-locked with a 
phase-shift, $\Delta\phi$, which can be computed by combining the expressions 
for $m_{osc}(t)$ and $Z(\phi)$ \cite{SuppMat}.
This $\Delta\phi$ depends on the local 
inhibition $K_0$ and on the delay $D$. 
We found that in general two regimes can be distinguished as a function of $K_0$
(see Fig.~\ref{fig:phasediagram}, inset).
In the first regime, $|K_0^{osc}(D)| < |K_0| < |K_{0}^{c}(D)|$, the activities of the two populations
oscillate in anti-phase ($\Delta \phi = \pi$).  
At $K_0 = K_0^c$ a supercritical pitchfork bifurcation occurs 
from anti-phase locking toward out-of-phase locking. 
For $|K_0| > |K_0^c(D)|$, two stable intermediate phase-shifts exist, 
$\Delta\phi = \pi \pm \xi, 0 < \xi <\pi$. 
Such dynamical configurations break the invariance of 
equations \eqref{RateEquation} under permutation of the the populations and a \textit{leader} population acquires a phase advance with
respect to a \textit{laggard} population (spontaneous symmetry breaking).
These two regimes of phase-locking persist if $K_1$ is not
too large (see Fig.~\ref{fig:phasediagram}).
However, when $K_1$ increases sufficiently, phase-locked oscillations 
destabilize and a series of bifurcations leads eventually to 
chaos. The largest Lyapunov exponent $\lambda_{max}$ ---evaluated 
by numerical integration of the linearized equations--- in fact becomes strictly 
positive (colored background in Fig.~\ref{fig:phasediagram}). 
The scenario for the onset of chaos depends critically on the strength 
of local inhibition as we show below.

\begin{figure}
\begin{center}
\includegraphics[scale=0.559,angle=-90]{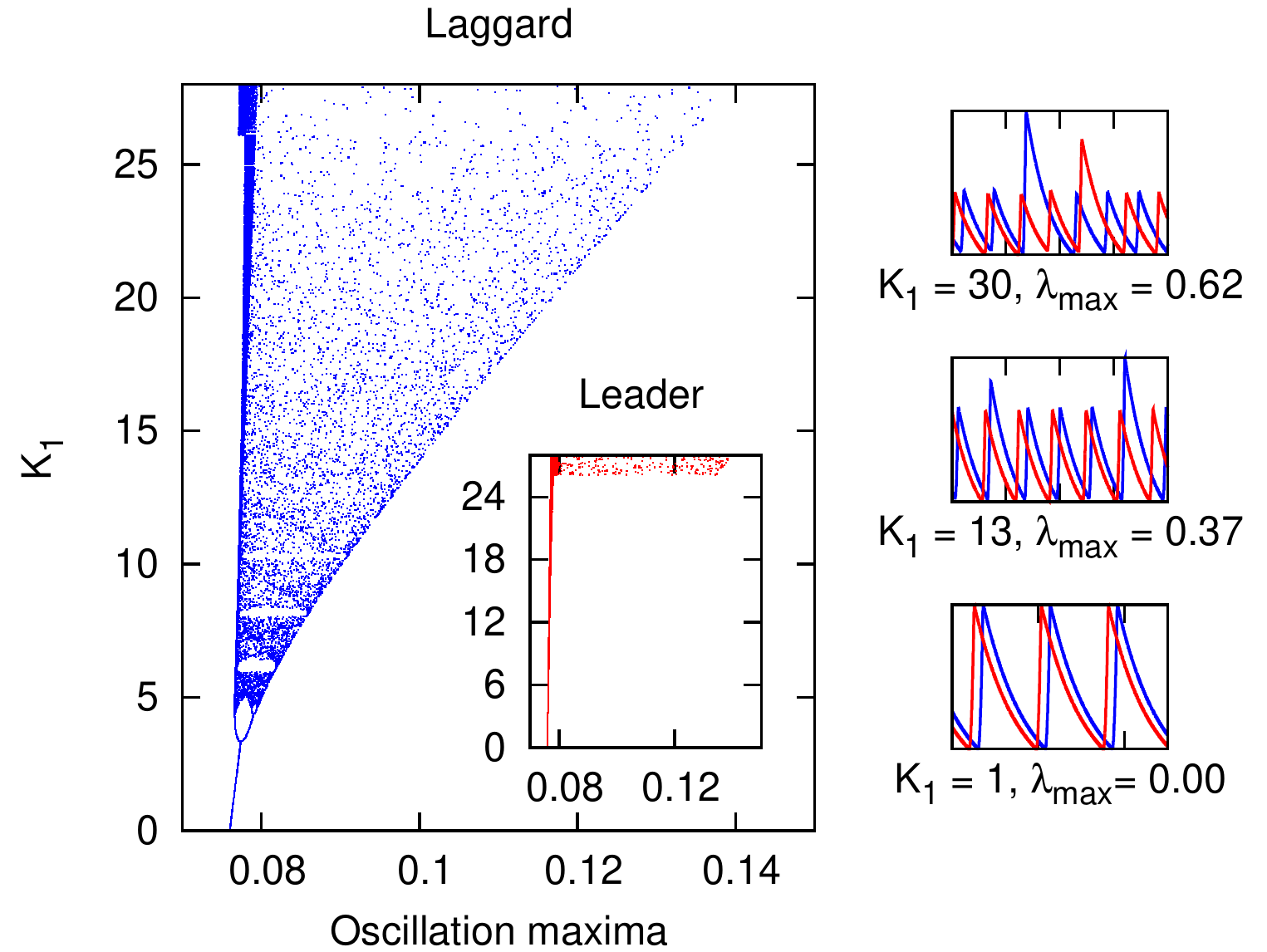}
\end{center}
\caption{Period-doubling scenario ($K_0 =-500$, $D=\bar D = 0.1$). Bifurcation diagram for the 
  laggard (blue) and for the leader populations (red). Side panels: activity traces ($\tau$ units). Bottom to top: out-of-phase ($K_1 = 1$), asymmetric ($K_1 = 13$) and symmetric chaos  ($K_1=30$).}\label{fig:bifo1}
\end{figure}
For sufficiently strong inhibition 
(region \textit{Feig.} in Fig.~\ref{fig:phasediagram}), 
chaos originates from out-of-phase locking via period doubling. This Feigenbaum 
scenario is  confirmed by the numerical estimations of the constants 
$\delta \simeq 4.66(9) \pm 0.002$ and $\alpha = 2.502(8) \pm 0.0001$, 
close to their universal values \cite{feig_original, SuppMat}.
The bifurcation diagram shown in  Fig.~\ref{fig:bifo1}
shows the Feigenbaum cascade and the chaotic regime for $K_0 = -500$ and $D=\bar{D}=0.1$. 
Dots in the figure and in the inset correspond to values of the activity at 
a local maxima for each population 
Remarkably, out-of-phase locking which occurs at 
weak-coupling gives rise to {\it asymmetric} chaos, for not too large $K_1$ ($K_1 \lesssim 27$). This can be seen in 
Fig.~\ref{fig:bifo1}, where fluctuations are considerably larger in the laggard population
than in the leader. The leader population oscillations are almost periodic
and can be thought of as driving the laggard. 
A very similar bifurcation sequence 
is found, indeed, if one takes the excitation between the two populations
to be unidirectional \cite{SuppMat}. As a matter of fact, it is well known that
such a system can exhibit chaotic behaviors 
\cite{synusentrainedchaos}. 
As $K_1$ increases further, an abrupt transition occurs to a symmetric chaotic state in which the fluctuations have 
comparable magnitudes in the two populations (see Fig.~\ref{fig:bifo1}). 
Eventually, the activity of both populations goes to infinity
(rate instability) when the positive feedback loop between 
them becomes exceedingly strong.

\begin{figure}
\begin{center}
\includegraphics[scale=0.559,angle=-90]{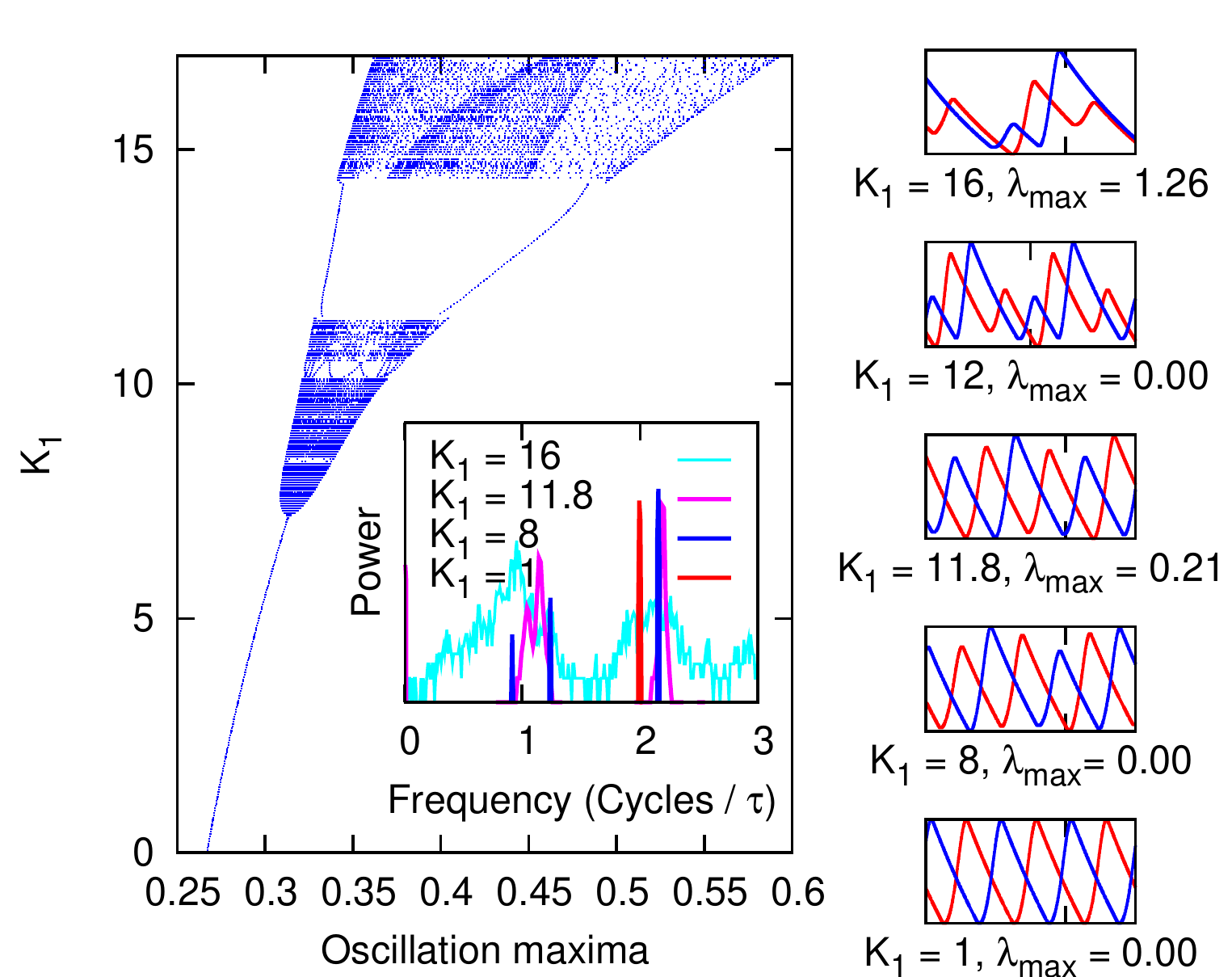}
\end{center}
\caption{Quasi-periodic scenario ($K_0 = -50, D=\bar D=0.1$).
 Bifurcation diagram. Inset: power spectra. Side panels: activity traces ($\tau$ units). Bottom to top:
 anti-phase ($K_1=1$), quasi-periodic ($K_1=8$), chaotic ($K_1=11.8$), doublets ($K_1=12$) and chaotic doublets ($K_1=16$).}\label{fig:bifo2}
\end{figure}
A different scenario occurs when considering the destabilization of the anti-phase locked periodic state (see Fig.~\ref{fig:phasediagram}).
The corresponding bifurcation
diagram is shown in Fig.~\ref{fig:bifo2}.
In this scenario no symmetry breaking occurs.
Quasi-periodic oscillations and eventually
chaotic oscillations emerge as $K_1$ 
increases. This is revealed by spectral analysis. As the excitation becomes stronger, two and then three incommensurate 
frequencies  appear in the power spectrum. The first 
occurrence of $\lambda_{max} >0$ is associated with the sudden broadening 
of the Fourier peaks (inset of Fig.~\ref{fig:bifo2}). This behavior 
corresponds to the Newhouse-Ruelle-Takens quasi-periodic scenario 
for the onset of chaos \cite{qp_original}. 
As shown in Figs.~\ref{fig:phasediagram} 
and~\ref{fig:bifo2}, chaos is intertwined with quasi-periodic 
and resonant windows of period-doubled regular oscillations (doublets). 

We conjecture that the dynamical properties described above
and the destruction of coherence by long range excitatory 
interactions are in fact a general feature of neuronal networks 
in which population synchronous oscillations are 
induced by local inhibition. 
We verified this claim in a large network model of conductance-based spiking neurons
consisting of two populations of Hodgkin-Huxley type neurons  
\cite{WangBuzsaki96}. Interactions among cells within a local population are 
purely inhibitory. For simplicity, these same cells are allowed to establish excitatory inter-population connections \cite{SuppMat}.

\begin{figure}
\begin{center}
\includegraphics[scale=0.526,angle=-90]{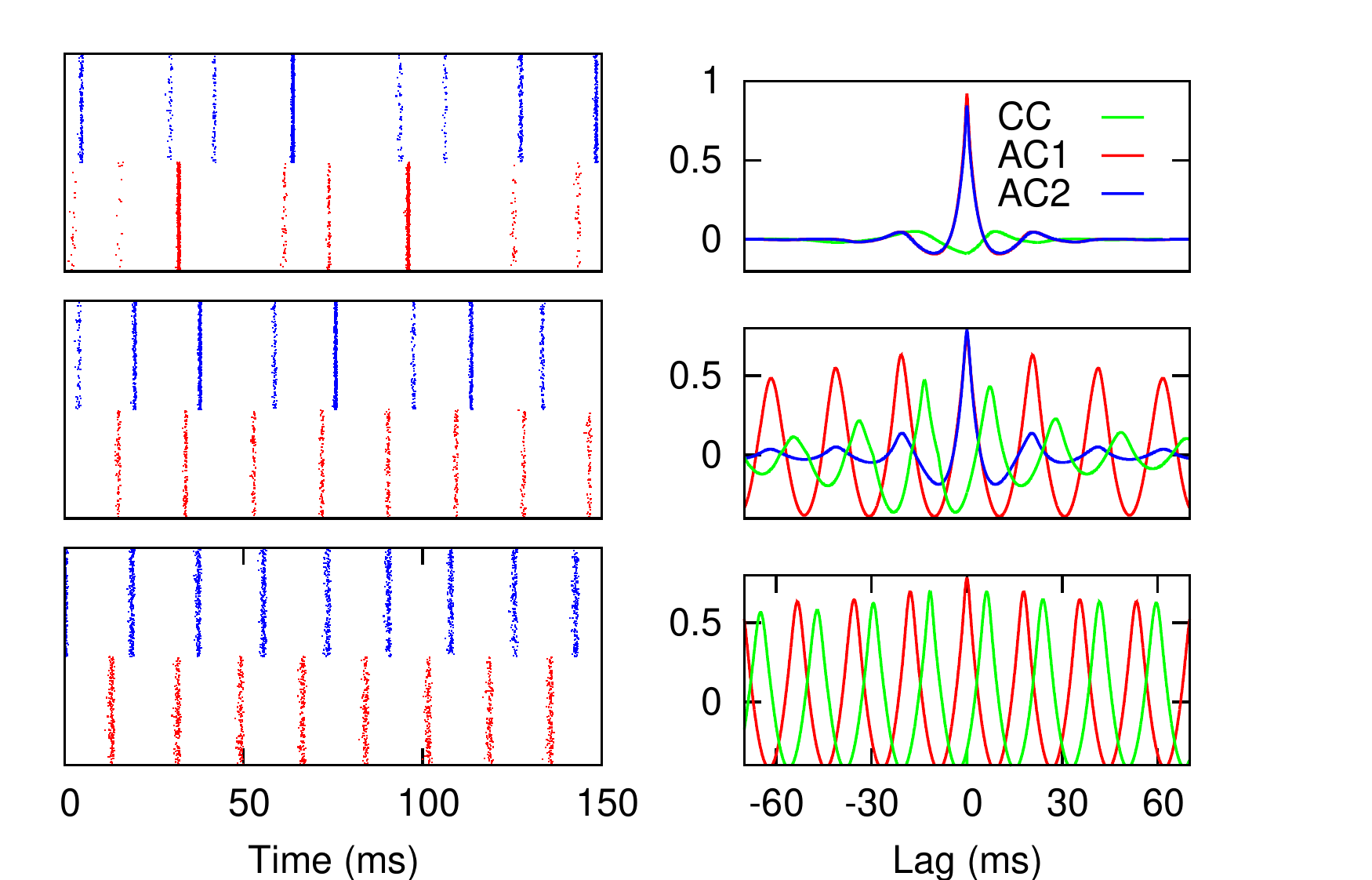}
\end{center}
\caption{Strong local inhibition (spiking neurons): $p^I=0.6$, 
$g^I=88 \,\mu$S/cm$^2$ (other parameters in \cite{SuppMat}). 
Left: raster plots (red, leader; blue, laggard population). Right: autocorrelograms (both shown for asymmetric states) and crosscorrelograms. Bottom to top: 
  out-of-phase ($p^E=0.01$),  
  asymmetrically ($p^E=0.10$) and symmetrically irregular ($p^E=0.14$). }
  \label{fig:condu1}
\end{figure}

\begin{figure}
\begin{center}
\includegraphics[scale=0.526,angle=-90]{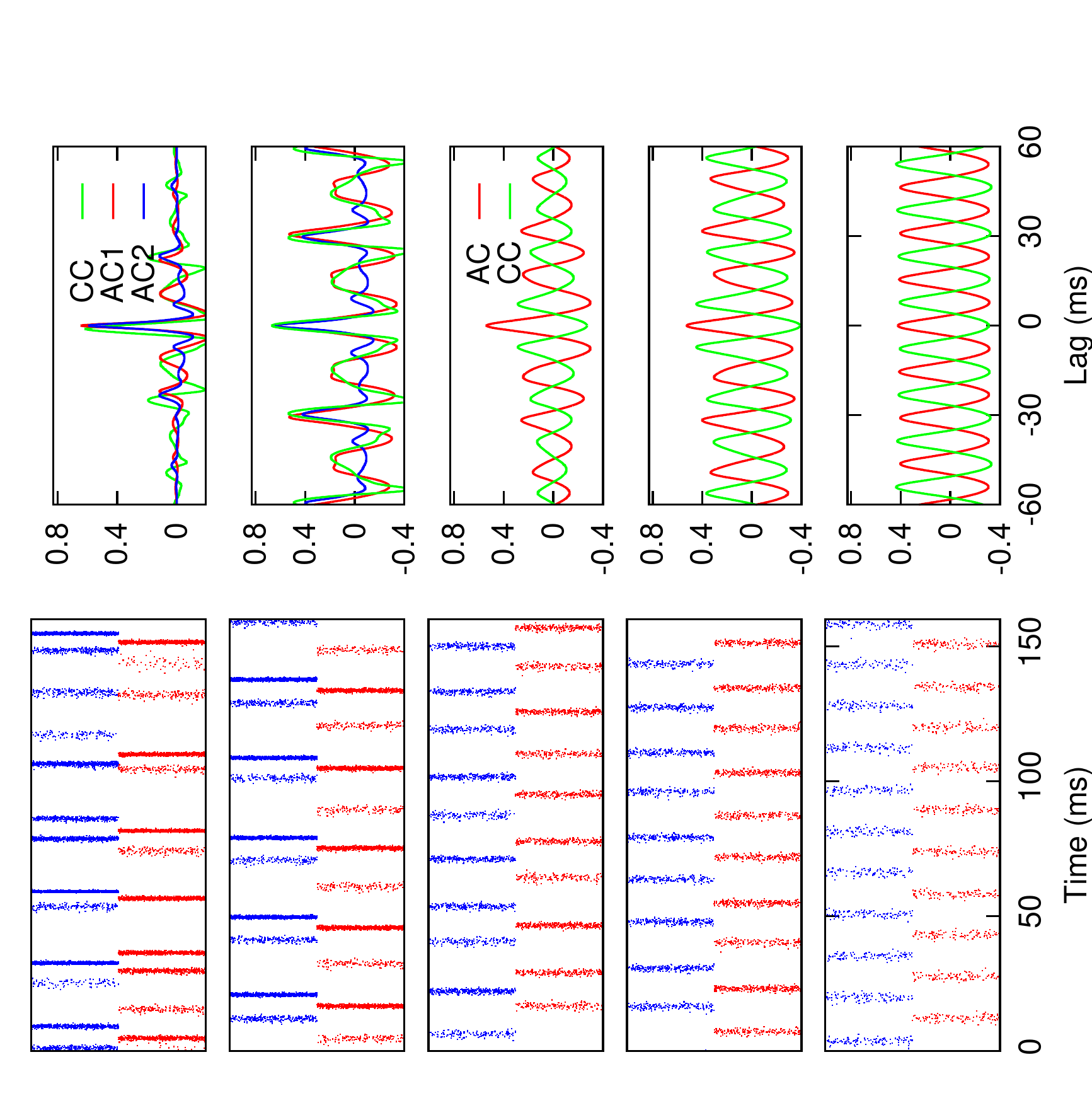}
\end{center}
\caption{Weak local inhibition (spiking neurons):  $p^I
=0.1$, $g^I=18 \,\mu$S/cm$^2$ (other parameters in \cite{SuppMat}). 
Left: raster plots (red, first; blue, second population). Right: autocorrelograms (both shown for asymmetric states) and crosscorrelograms. 
Bottom to top:  anti-phase ($p^E=0.008$);  quasi-periodic ($p^E=0.017$); irregular ($p^E=0.032$); 
  doublets ($p^E =0.035$); irregular doublets ($p^E=0.062$).}\label{fig:condu2}
\end{figure}

The connectivity patterns are random with a probability of connection $p^I$
(resp. $p^E$) for  two neurons in the same (resp. different) populations. 
Synaptic couplings are modeled as time-varying conductances 
(peak conductances $g^{I,E}$, rise and decay time $\tau_1$ and $\tau_2$, delay $d$ \cite{SuppMat}). The parameters of the network and 
of the external tonic input are fixed in order to obtain fast 
synchronous oscillations in the gamma frequency band when 
$p^E=0$ \cite{SuppMat}. 
The strength of the inter-population excitation 
is then modulated by varying $p^{E}$. 
The temporal decorrelation of the oscillations and the phase relation 
between the population activities are characterized by 
the autocorrelograms (ACs) and the crosscorrelogram (CC) of 
the average neuronal voltages 
$\langle V^{(\alpha)}\rangle = \frac{1}{N}\sum_j V_j^{(\alpha)}$,
where $V_j^{(\alpha)}$ represents the voltage of the $j-$th neuron ($j=1,\ldots, N$)
in population $\alpha=1,2$.
Results for $N=32000$ are shown in Fig.~\ref{fig:condu1} and in
Fig.~\ref{fig:condu2} for a strong and relatively weak local inhibition, 
respectively. 

In the case of strong local inhibition,  out-of-phase oscillations 
are observed when $p^E$ is small (see CCs in the right column in Fig.~\ref{fig:condu1}). 
For increasing  $p^E$ the oscillations gradually become more
irregular, at first only 
in the laggard population and then in both populations. As a result, 
the ACs of $\langle V \rangle$ become rapidly damped with 
decorrelation times on the order of tens of milliseconds (right column of 
Fig.~\ref{fig:condu1}, laggard in blue, leader in red). 
Note that a high degree of synchrony within each population is maintained even 
when the oscillations are irregular (ACs are normalized 
$\chi=(\sigma^2_{\langle V \rangle}/\langle \sigma^2_{V_j}\rangle)^\frac{1}{2}$  \cite{GolombHansel00}).  

If the local inhibition is not too strong, the oscillations 
of the two populations lock in anti-phase for small $p^E$. 
They gradually become more irregular when $p^E$ is increased,
but the fluctuations are now similar
in the two populations. 
Characteristic modulations in the envelope of the ACs hint 
at a quasi-periodic scenario for the emergence of irregular activity (see Fig.~\ref{fig:condu2}). 

The sequences
of raster plots displayed for increasing $p^E$ in
Figs.~\ref{fig:condu1} and ~\ref{fig:condu2} reproduce the main
qualitative features of the corresponding Figs. ~\ref{fig:bifo1} and
~\ref{fig:bifo2}. Finally, a more realistic network
in which each population contains separate excitatory and inhibitory cells,
and in which excitatory cells fire at lower rates than inhibitory cells,
exhibit qualitatively similar dynamics (see \cite{SuppMat}).

Our combined analytical and numerical study therefore suggests that
large scale networks in which local interactions are predominantly inhibitory
can exhibit two robust routes to synchronized chaotic states as the
strength of long-range excitation is increased. In
our model, oscillations are generated locally through delayed
inhibition, leading to a stochastic
synchronized state in which neurons fire typically less than one spike
per cycle of the oscillations \cite{Brunel99}. Our scenario is
consistent with experimental observations {\em in vivo}
\cite{TukkerKlausberger07}. However, other scenarios have been proposed in
which gamma oscillations emerge due to spike-to-spike synchrony, driven by mutual inhibition, or to the excitation-inhibition feedback loop (see e.g. \cite{WhittingtonKopell00, Mato01}). Further
experimental studies are necessary to elucidate the mechanisms
underlying the generation of gamma oscillations {\em in vivo}.

Previous modeling studies have considered the role of long-range
excitation in synchronizing the activity of distant neuronal
assemblies \cite{borisyuk, Traub96, Kopell98}.  Here, we have found that
chaotic activity naturally emerges when the long-range excitation is
sufficiently strong.  Thus, the locally generated rhythmic
activity undergoes temporal decorrelation, even though, 
at zero time-lag, 
the degree of
synchronization 
between the populations increases for larger excitatory coupling.
Such tightly synchronized firing might be an effective
way to drive the connectivity between these populations through synaptic plasticity \cite{BiPoo98}. 
Besides, effective long range interactions between populations of neurons in
primary visual cortex may be modulated by the spatial patterns of a
visual stimulus \cite{grannan_kleinf_sompo_93}. Our work predicts then stimulus-dependent decoherence on synchronous activity evoked in the visual cortex. 
 
This research was funded by the UniNet EU excellence network, the NeuralComp initiative and the Franco-Israeli Laboratory of Neurophysics and System Neurophysiology.

\end{document}